\documentclass[12pt]{iopart}
\usepackage{epsfig}
\usepackage{graphicx}
\usepackage{color}
\eqnobysec
\begin{document}

\title{Statefinder diagnostic in a torsion cosmology}

\author{Xin-zhou Li, Chang-bo Sun and Ping Xi}

\address{Shanghai United Center for Astrophysics(SUCA),
 Shanghai Normal University, 100 Guilin Road, Shanghai 200234,China}
\ead{kychz@shnu.edu.cn}
\begin{abstract}
We apply the statefinder diagnostic to the torsion cosmology, in
which an accounting for the accelerated universe is considered in
term of a Riemann-Cartan geometry: dynamic scalar torsion. We find
that there are some typical characteristic of the evolution of
statefinder parameters for the torsion cosmology that can be
distinguished from the other cosmological models. Furthermore, we
also show that statefinder diagnostic has a direct bearing on the
critical points. The statefinder diagnostic divides the torsion
parameter $a_1$ into differential ranges, which is in keeping with
the requirement of dynamical analysis. In addition, we fit the
scalar torsion model to ESSENCE supernovae data and give the best
fit values of the model parameters.
\end{abstract}

\maketitle

\section{Introduction}
The current observations, such as SNeIa (Supernovae type Ia), CMB
(Cosmic Microwave Background) and large scale structure, converge on
the fact that a spatially homogeneous and gravitationally repulsive
energy component, referred as dark energy, accounts for about $70$
\% of the energy density of universe. Some heuristic models that
roughly describe the observable consequences of dark energy were
proposed in recent years, a number of them stemming from a certain
physics \cite{r1} and the others being purely phenomenological
\cite{r2}. Dark energy can even behave as a phantom and effectively
violate the weak energy condition\cite{phantom}.

In various cosmological models, fundamental quantities are either
geometrical (if they are constructed from a spacetime geometry
directly) or physical (if they depend upon physical fields).
Physical quantities are certainly model-dependent, while geometrical
quantites are more universal. About thirty years ago, the bouncing
cosmological model with torsion was suggested in Ref.\cite{Kerlick},
but the torsion was imagined as playing role only at high densities
in the early universe. Goenner et al. made a general survey of the
torsion cosmology \cite{Goenner}, in which the equations for all the
PGT (Poincar{\'e} Gauge Theory of gravity) cases were discussed
although they only solved in detail a few particular cases. Recently
some authors have begun to study torsion as a possible reason of the
accelerating universe \cite{Boeheretal}. Nester and collaborators
\cite{r3} consider an accounting for the accelerated universe in
term of a Riemann-Cartan geometry: dynamic scalar torsion. They
explore the possibility that the dynamic PGT connection, reflecting
the nature of dynamic PGT torsion, provides the accelerating force.
With the usual assumptions of homogeneity and isotropy in cosmology
and specific cases of the suitable parameters and initial
conditions, they find that torsion field could play a role of dark
energy.

One of the motivation was to avoid singularities in the initial
investigations of torsion cosmology \cite{b1}. However, it soon was
found that non-linear torsion effects were more likely to produce
stronger singularities \cite{b2}. The non-linear effects turn out to
play a key role for the outstanding present day mystery: the
accelerated universe. In the various PGT, the connection dynamics
decomposed into six modes with certain spin and parity: $2^\pm$,
$1^\pm$, $0^\pm$. Some investigations showed that $0^\pm$ may well
be the only acceptable dynamic PGT torsion modes \cite{b3}. The
pseudoscalar mode $0^-$ is naturally driven by the intrinsic spin of
elementary fermions, therefore it naturally interacts with such
sources. Consequently, it is generally thought that axial torsion
must be small and have small effects at the late time of
cosmological evolution. This is a major reason why one does not
focus on this mode at the late time. On the other hand, the scalar
mode $0^+$ does not interact in any direct obvious fashion with any
known type of matter \cite{b4}, therefore one can imagine it as
having significant magnitude and yet not being conspicuously
noticed. Furthermore, there is a critical non-zero value for the
affine scalar curvature since $0^+$ mode can interact indirectly
through the non-linear equations. The homogeneity and isotropy of
cosmology have received strong confirmation from modern
observations, which greatly restrict the possible types of
non-vanishing fields. Under the assumption of homogeneity and
isotropy, $0^+$ mode has only a time component and it can be
specified as the gradient of a time-dependent function. Therefore,
the cosmological models with the scalar mode offer a situation where
dynamic torsion may lead to observable effect at late time. We
emphasize again that one does not focus on the early universe, where
one could indeed expect large effects (though their signature would
have to be separated from other large effects), and substitutionally
asks about traces of torsion effects at the late time of
cosmological evolution \cite{r3}.

Obviously, the fine-tuning problem is one of the most important
issues for the torsion cosmology \cite{r3}. And a good model should
limit the fine-tuning as much as possible. The dynamical attractor
of the cosmological system has been employed to make the later time
behaviors of the model insensitive to the initial condition of the
field and thus alleviates the fine-tuning problem \cite{r4}.
Furthermore, Nester et al \cite{r3} have shown that the Hubble
parameter and $\ddot{a}$ have an oscillatory form for the scalar
torsion cosmology.

The traditional geometrical parameters, i.e., the Hubble parameter
$H \equiv \dot{a}/a$ and the deceleration parameter $q \equiv
-\ddot{a}a/\dot{a}^2$, are two elegant choices to describe the
expansion state of universe but they can not distinguish various
accelerating mechanism uniquely, because a quite number of models
may just correspond to the same current values of $H$ and $q$.
However, Sahni, Saini, Starobinsky and Alam \cite{r5} have
introduced the statefinder pair $\{r, s\}$: $r \equiv
\stackrel{\dots}{a}/aH^3$, $s \equiv (r-1)/3(q-1/2)$. It is
obviously a natural next step beyond $H$ and $q$. Fortunately, as is
shown in the literatures \cite{r6,r9}, the statefinder parameters
which are also geometrical diagnostics, are able to differentiate a
series of cosmological models successfully. Using the discussion of
statefinder parameters in the scalar torsion cosmology, we explain
easily why the present field equations modify the expansion of the
universe only at late time. If the evolving trajectory of
statefinder have a decelerating phase ($q > 0$) at early time, then
we can understand why the expansion of the universe until $z \sim
few$ remains unchanged in the scalar torsion models.

In this paper, we apply the statefinder diagnostics to the torsion
cosmology. We find that there are some characteristics of
statefinder parameters for the torsion cosmology that can be
distinguished from the other cosmological models. The statefinder
diagnostics show that the universe naturally has an accelerating
expansion at low redshifts (late time) and a decelerating expansion
at high redshifts (early time). Therefore, scalar torsion cosmology
can avoid some of the problems which occur in other models.
Especially, the effect of torsion can make the expansion rate
oscillate when torsion parameter $a_1 > 0$ or $a_1 < -1$. Whether
the universe has properties which are easier to explain within the
scalar torsion context is a remarkable possibility demanding further
exploration. The oscillatory feature of Hubble parameter had earlier
been reported for the braneworld cosmology \cite{P} and the
quasi-steady state cosmology \cite{F}. We show that statefinder
diagnostic has a direct bearing on the critical points of the
dynamical system. One of the most interesting characteristic of the
trajectories is that there are loop and curves with the shape of
tadpole in the case of the torsion parameter $-\frac{1}{9} \leq a_1
< 0$. In this case, we fit the scalar torsion model to current type
Ia supernova data and find it is consistent with the observations.
Furthermore, we analyze preliminarily the relevance for realistic
observation of the found statefinder parameters.

\section{The equations of motion}
PGT \cite{r7} has been regarded as an interesting alternative to
general relativity because of its gauge structure and geometric
properties. PGT based on a Riemann-Cartan geometry, allows for
dynamic torsion in addition to curvature. The affine connection of
the Riemann-Cartan geometry is
\begin{equation}
  \Gamma_{\mu\nu}{}^\kappa=\overline{\Gamma}_{\mu\nu}{}^\kappa+\frac{1}{2}(T_{\mu\nu}{}^\kappa+T^\kappa{}_{\mu\nu}
  +T^\kappa{}_{\nu\mu})\,\label{PGT}
\end{equation}
where $\overline{\Gamma}^{\kappa}_{\mu\nu}$ is the Levi-Civita
connection and $T_{\mu\nu}^{\kappa}$ is the torsion tensor.
Meantime, the Ricci curvature and scalar curvature can be written as
\begin{eqnarray}
   R_{\mu\nu} &=& \overline{R}_{\mu\nu} + \overline{\nabla}_\nu T_\mu +\frac{1}{2}
   (\overline{\nabla}_\kappa - T_\kappa)(T_{\nu\mu}{}^\kappa+T^\kappa{}_{\mu\nu}+T^\kappa{}_{\nu\mu})\nonumber\\
   &&+\frac{1}{4}(T_{\kappa\sigma\mu}T^{\kappa\sigma}{}_\nu+2T_{\nu \kappa \sigma}T^{\sigma \kappa}{}_\mu)\,,\\
   R&=&\overline{R} + 2\overline{\nabla}_\mu T^\mu+\frac{1}{4}(T_{\mu\nu \kappa}T^{\mu\nu \kappa}
    +2T_{\mu\nu \kappa}T^{\kappa\nu\mu}-4T_\mu T^\mu)\,,\label{RiSCur}
\end{eqnarray}
where $\bar{R}_{\mu\nu}$ and $\bar{R}$ are the Riemannian Ricci
curvature and scalar curvature, respectively, and $\bar{\nabla}$ is
the covariant derivative with the Levi-Civita connection (For a
detailed discussion see Ref. \cite{r3}.). Theoretical analysis of
PGT led us to consider tendentiously dynamic "scalar mode". In this
case, the restricted expression of the torsion can be written as
\cite{r3}
\begin{equation}
  T_{\mu\nu\kappa}=\frac{2}{3}T_{[\mu}g_{\nu]\kappa},\label{Trestrct}
\end{equation}
where the vector $T_\mu \equiv T_{\mu\nu}^{\nu}$ is the trace of the
torsion. Then, the Ricci curvature and scalar curvature can be
expressed as
\begin{eqnarray}
    R_{\mu\nu}&=&\overline{R}_{\mu\nu}+\frac{1}{3}(2\overline{\nabla}_\nu T_\mu+g_{\mu\nu}
   \overline{\nabla}_kT^k)+\frac{2}{9}(T_\mu T_\nu-g_{\mu\nu}T_\kappa T^\kappa)\,\\
   R&=&\overline{R} + 2\overline{\nabla}_\mu T^\mu-\frac{2}{3}T_\mu T^\mu\,.\label{RiSCurrestrict}
\end{eqnarray}
 The gravitational Lagrangian density for the scalar mode
is
\begin{eqnarray}
  L&=& -\frac{a_0}{2}R +\frac{b}{24}R^2
       +\frac{a_1}{8}(T_{\nu\sigma\mu}T^{\nu\sigma\mu}
         +2T_{\nu\sigma\mu}T^{\mu\sigma\nu}-4T_\mu
         T^\mu)\,,\label{Lg0+mode}
\end{eqnarray}
where $R\equiv R_{\mu\nu}^{\mu\nu}$ and $a_1$ is a torsion
parameter. Consider that the parameter $b$ is associated with
quadratic scalar curvature term $R^2$, so that $b$ should be
positive \cite{r3}. Therefore, the field equations of the scalar
mode are
\begin{eqnarray}
   &&\overline{\nabla}_\mu R +\frac{2}{3}(R+\frac{6\mu}{b})T_\mu=0\,,\label{gradR}\\
  &&a_0(\overline{R}_{\mu\nu}-\frac{1}{2}g_{\mu\nu}\overline{R})
  =-({\cal T}_{\mu\nu}+\widetilde{\cal T}_{\mu\nu}),\label{fe1}
\end{eqnarray}
where ${\cal T}_{\mu\nu}$ is the source energy-momentum tensor and
$\widetilde{\cal T}_{\mu\nu}$ is the contribution of the scalar
torsion mode to the effective total energy-momentum tensor:
\begin{eqnarray}
   \widetilde{\cal T}_{\mu\nu}=&&-\frac{2\mu}{3}(\overline{\nabla}_{(\mu}T_{\nu)}
  -g_{\mu\nu}\overline{\nabla}_\kappa T^\kappa)-\frac{\mu}{9}(2T_\mu T_\nu+g_{\mu\nu}T_\kappa
  T^\kappa)\nonumber\\
    &&-\frac{b}{6}R(R_{(\mu\nu)}-\frac{1}{4}g_{\mu\nu}R)\,.\label{torT}
\end{eqnarray}

Since current observations favor a flat universe, we will work in
the spatially flat Robertson-Walker metric
$ds^2=-dt^2+a^2(t)[dr^2+r^2(d\theta^2+\sin^2\theta d\phi^2)]$, where
$a(t)$ is the scalar factor. This engenders the Riemannian Ricci
curvature and scalar curvature:
\begin{eqnarray}
  &&\overline{R}_t{}^t = 3\frac{\ddot{a}}{a} = 3(\dot{H} + H^2)\,,\\
  &&\overline{R}_r{}^r = \overline{R}_\theta{}^\theta =\overline{R}_\phi{}^\phi
                  =\frac{\ddot{a}}{a} + 2\frac{\dot{a}^2}{a^2}
                  = \dot{H} + 3H^2\,,\\
  &&\overline{R}=6(\frac{\ddot{a}}{a}+\frac{\dot{a}^2}{a^2})=6(\dot{H} + 2H^2)\,,
\end{eqnarray}
where $a(t)$ is the scalar factor, and $H$ is the Hubble parameter.
The torsion $T_\mu$ should also be only time dependent, therefore
one can let $T_{t}(t)\equiv \Phi(t)$ ($\Phi(t)$ is the torsion
field) and the spatial parts vanish. The corresponding equations of
motion in the matter-dominated era are as follows
\begin{eqnarray}
      \dot{H}&=&\frac{\mu}{6a_1}R-\frac{\rho}{6a_1}-2H^2\,,\label{dtH}\\
      \dot{\Phi}&=&-\frac{a_0}{2a_1}R-\frac{\rho}{2a_1}-3H\Phi
                   +\frac{1}{3}\Phi^2\,,\label{dtphi}\\
      \dot{R}&=&-\frac{2}{3}\left(R+\frac{6\mu}{b}\right)\Phi\,,\label{dtR}
\end{eqnarray}
where $\mu= a_1-a_0$ and the energy density of matter component
\begin{eqnarray}
  &&\rho=\frac{b}{18}(R+\frac{6\mu}{b})(3H-\Phi)^2-\frac{b}{24}R^2-3a_1H^2
  \,.\label{fieldrho}
 \end{eqnarray}

One can scale the variables and the parameters as
\begin{eqnarray}
&&t\rightarrow l_{p}^{-2}H_{0}^{-1}t,\,\, H\rightarrow
l_{p}^{2}H_{0} H,
\,\, \Phi\rightarrow l_{p}^{2}H_{0}\Phi,\,\, R\rightarrow l_{p}^{4}H_{0}^{2}R,\nonumber\\
&&a_0\rightarrow l_{p}^{2}a_0,\,\, a_1\rightarrow l_{p}^{2}a_1,\,\,
\mu\rightarrow l_{p}^{2}\mu,\,\, b\rightarrow
l_{p}^{-2}H_{0}^{-2}b,\label{scale}
\end{eqnarray}
where $H_0$ is the present value of Hubble parameter and
$l_p\equiv\sqrt{8\pi G}$ is the Planck length. Under the transform
(\ref{scale}), Eqs. (\ref{dtH})-(\ref{dtR}) remain unchanged. After
transform, new variables $t$, $H$, $\Phi$ and $R$, and new
parameters $a_0$, $a_1$, $\mu$ and $b$ are all dimensionless.
Obviously, the Newtonian limit requires $a_0=-1$.

For the case of scalar torsion mode, the effective energy-momentum
tensor can be represented as
\begin{eqnarray}
\widetilde{T}_t{}^t&=&\!\!-3\mu H^2\!+\frac{b}{18}(R+\frac{6\mu}{b})
  (3H-\Phi)^2-\frac{b}{24}R^2,\label{torho}\\
   \widetilde{T}_r{}^r&=&\widetilde{T}_\theta{}^\theta
   =\widetilde{T}_\phi{}^\phi
     ={1\over 3}[\mu(R-\overline{R})-\widetilde{T}_t{}^t]\,,\label{torpre}
\end{eqnarray}
and the off-diagonal terms vanish. The effective energy density
\begin{eqnarray}
\rho_{eff}=\rho+\rho_{T}\equiv
\rho+\widetilde{T}_{tt},\label{effrho}
  \end{eqnarray}
which is deduced from general relativity. $p_{eff}=p_T\equiv
\tilde{T}_r^r$ is an effective pressure, and the effective equation
of state is
\begin{eqnarray}
w_{eff}=\frac{\widetilde{T}_{r}^{r}}{\rho+\widetilde{T}_{tt}}\label{effw}
  \end{eqnarray}
which is induced by the dynamic torsion.

\begin{figure}[!htbp]
\includegraphics[width=7cm]{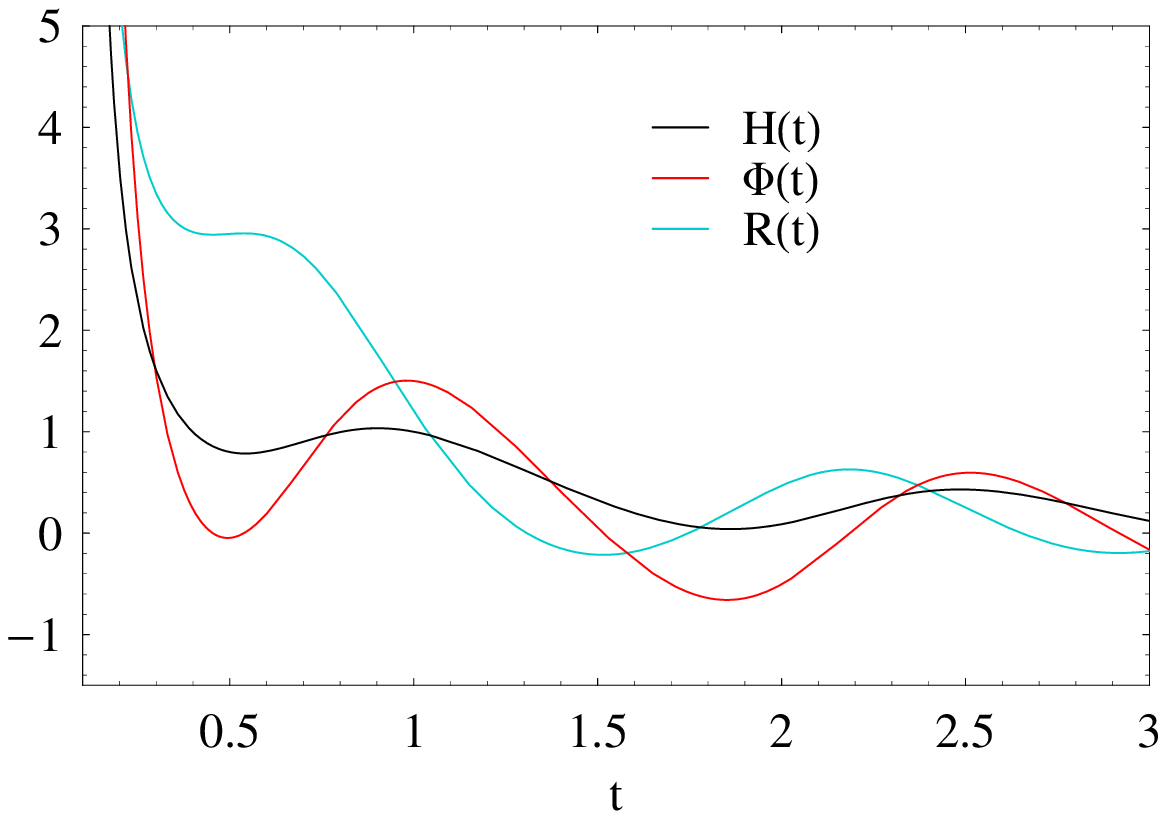}
\includegraphics[width=7.1cm]{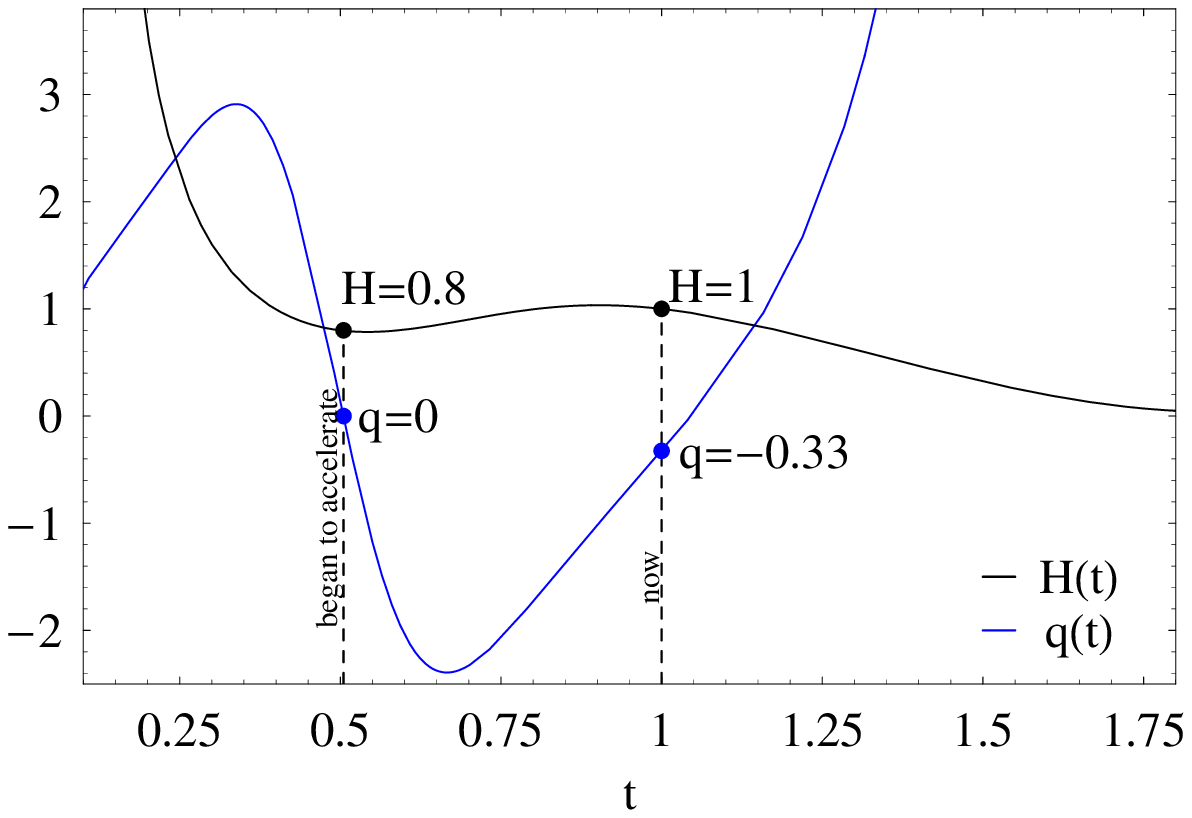}
\caption{Evolution of the Hubble parameter $H$, the temporal
component of the torsion $\Phi$, the affine scalar curvature $R$ and
the deceleration parameter $q$ as functions of time. We have chosen
the parameters $a_{1}=0.07$, $b=1.50$ and the initial values
$H(1)=1$, $\Phi(1)=1.50$, $R(1)=1.20$.}
 \label{Hqevol}
\end{figure}

In the case of $a_{1}>0$, Nester et al showed that the scalar mode
can contribute an oscillating aspect to the expansion rate of the
universe \cite{r3}. This oscillatory nature can be illustrated in
Fig. \ref{Hqevol} where we have chosen $a_{1}=0.07$, $b=1.50$,
$H(1)=1$, $\Phi(1)=1.50$, $R(1)=1.20$ and set the current time
$t_0=1$. According to scaling (\ref{scale}), the present value of
the Hubble parameter is unity. Obviously, $H$ is damped-oscillating
at late time and $q=-0.33$ is negative today, which means the
expansion of the universe is currently accelerating. The value of
$q$ turns from positive to negative when the time is around
$t\approx 0.51$, which is the epoch the universe began to
accelerate.

However, the above result is dependent on the choice of initial data
and the values of the parameters. Then, the scalar torsion cosmology
is unsuited to solving the fine-tuning problem in the case of
$a_{1}>0$. In the following sections, we'll investigate the
statefinder and give the dynamics analysis for all ranges of the
parameter $a_1$.

\section{Statefinder diagnostic}
For the spatially flat $\Lambda$CDM model the statefinder parameters
correspond to a fixed point $\{1, 0\}$ while $\{1, 1\}$ for the
standard cold dark matter model (SCDM) containing no radiation.
Since the torsion cosmology have used the dynamic scalar torsion (a
geometry quantity in the Riemann-Cartan spacetime), the torsion
accelerating mechanism is bound to exhibit an essential distinction
in contrast with various dark energy models. Therefore, its
statefinder diagnostic is sure to reveal differential feature. Let
us now study the torsion cosmological model in detail. Using Eqs.
(\ref{dtH})-(\ref{fieldrho}), we have the deceleration parameter
\begin{eqnarray}
q=\frac{1}{2}+\left[4(6\mu +bR)(\Phi -3H)^{2}-3R(24\mu
+bR)\right](432a_{1}H^{2})^{-1}\,,\label{torsionq}
  \end{eqnarray}
and the statefinder parameters
\begin{eqnarray}
r=1+(6\mu
+bR)\left[bH(36H^{2}-3R+4\Phi^{2}-24H)-12\Phi\right](108a_{1}bH^{3})^{-1}\,,\nonumber\\
\label{torsionr}
  \end{eqnarray}
and
\begin{eqnarray}
s=\frac{4(6\mu+bR)\left[4bH(3H-\Phi)^{2}-3bHR-12\mu
\Phi\right]}{3bH\left[(6\mu +bR)(36H^{2}-24H\Phi
+4\Phi^{2}-3R)-54\mu R\right]}\,. \label{torsions}
  \end{eqnarray}

\begin{figure}[!htbp]
\includegraphics[width=7cm]{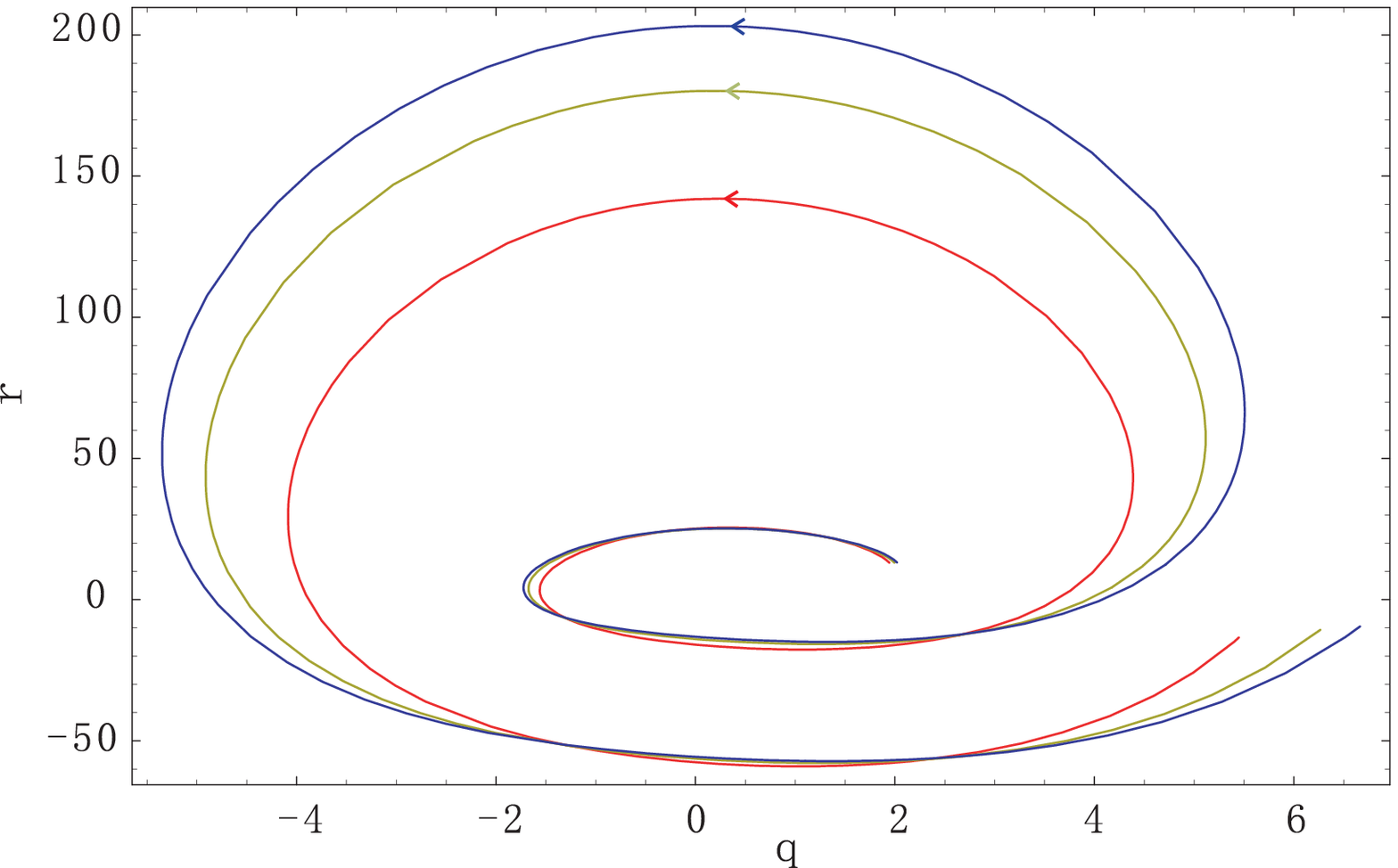}
\includegraphics[width=7cm]{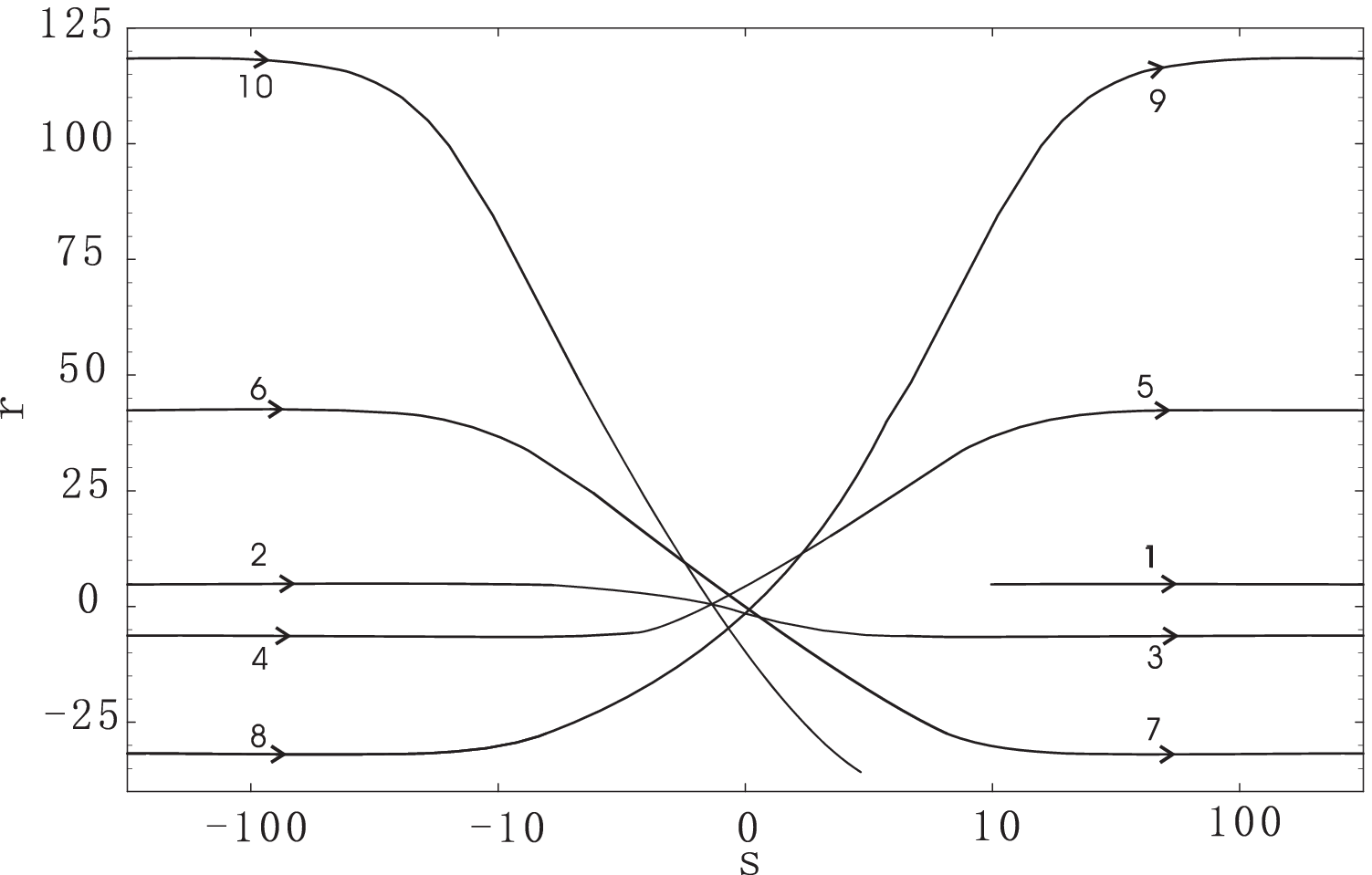}
\caption{Evolving trajectories of the statefinder in the $q-r$ and
$s-r$ planes for the case of torsion parameter $a_{1}\geq 0$, where
we choose the parameters $a_{1}=\frac{1}{2}$ and $b=4$. The arrows
show the direction of the time evolution.}
 \label{planecase4}
\end{figure}

In the following we will discuss the statefinder for four
differential ranges of the torsion parameter $a_1$, respectively.
Firstly, we consider the time evolution of the statefinder pairs
$\{r, s\}$ and $\{r, q\}$ in the case of $a_1 \geq 0$. In Fig.
\ref{planecase4}, we plot evolution trajectories in the $q-r$ and
$s-r$ planes, where we have chosen $a_1 = 1/2$ and $b = 4$. We see
easily that cosmic expansion alternates between deceleration and
acceleration in the evolving trajectories of $q-r$ plane, and the
amplitude becomes larger and larger as increase of time. The
trajectories in the $s-r$ plane is quite complicated, so we mark its
sequence by the ordinal number. Every odd number curve evolves from
finite to infinite, but even number curve evolves from infinite to
finite. These are quasi-periodic behaviors which corresponds to the
numerical solution of Ref. \cite{r3}. Noticeably, the trajectories
will never pass $\Lambda$CDM point $\{1, 0\}$.

\begin{figure}[!htbp]
\includegraphics[width=7cm]{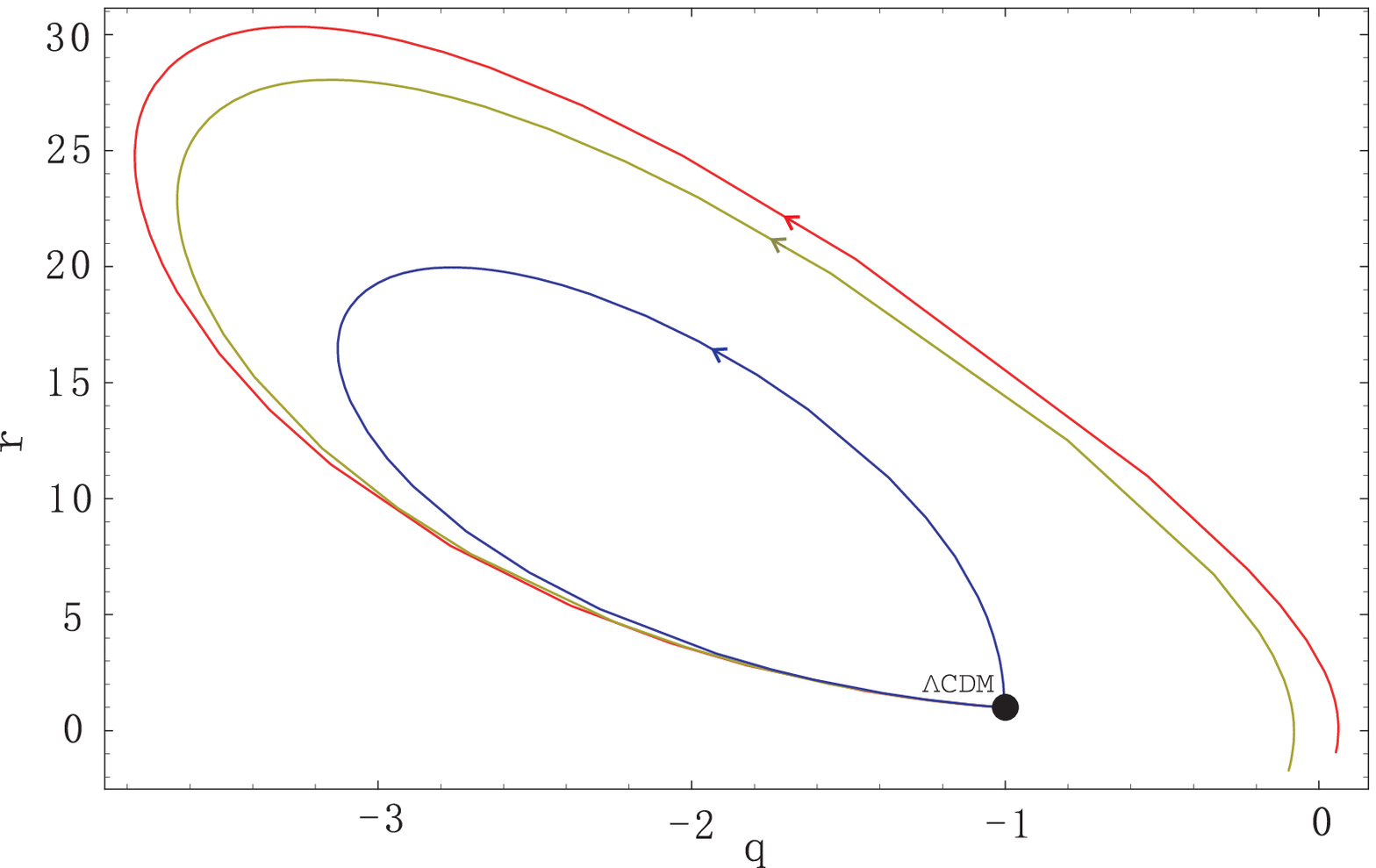}
\includegraphics[width=7cm]{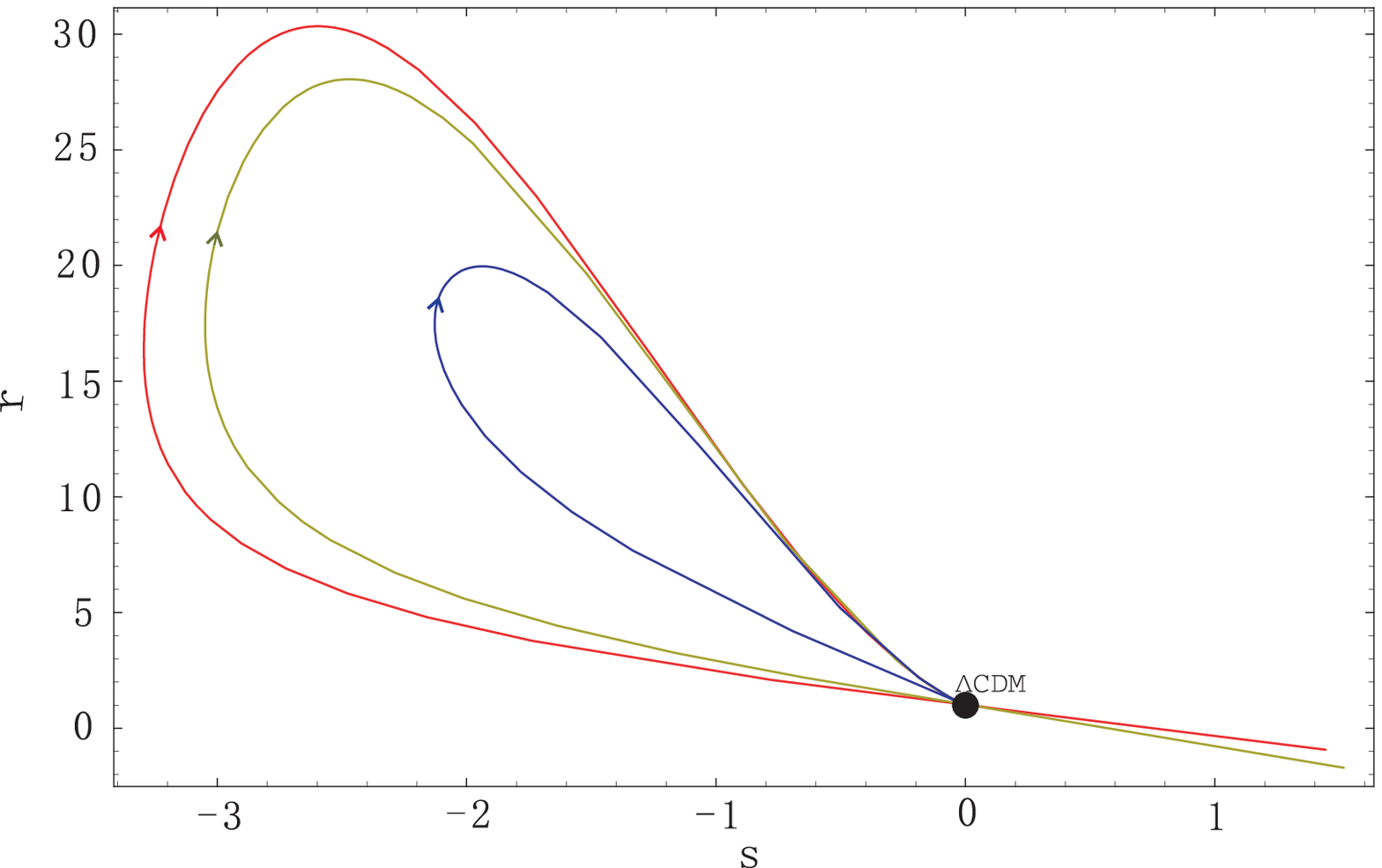}
\caption{Evolving trajectories of the statefinder in the $q-r$ and
$s-r$ planes for the case of torsion parameter $-\frac{1}{9}\leq
a_{1}<0$, where we choose the parameters $a_{1}=-\frac{1}{10}$ and
$b=3$. The arrows show the direction of the time evolution.}
 \label{planecase3}
\end{figure}

Secondly, we discuss the time evolution of the statefinder pairs
$\{r, s\}$ and $\{r, q\}$ for the case of $-\frac{1}{9} \leq a_1 <
0$. We plot evolving trajectories in Fig. \ref{planecase3}, where we
have chosen $a_1 = -\frac{1}{10}$ and $b = 3$. We see clearly that
the cosmic acceleration is guaranteed by the dynamic scalar torsion
in the evolving trajectories of $q-r$ plane, and the curves will
converge into $\Lambda$CDM point. The evolving trajectories go
through a climbing-up stage first, then get into a rolling-down
stage in the $s-r$ plane. Lastly, trajectories tend to $\Lambda$CDM
point $\{1, 0\}$. Furthermore, the only one forms a loop that starts
from $\{1, 0\}$ then evolves back to $\{1, 0\}$, and others show in
the shape of tadpole.

\begin{figure}[!htbp]
\includegraphics[width=7cm]{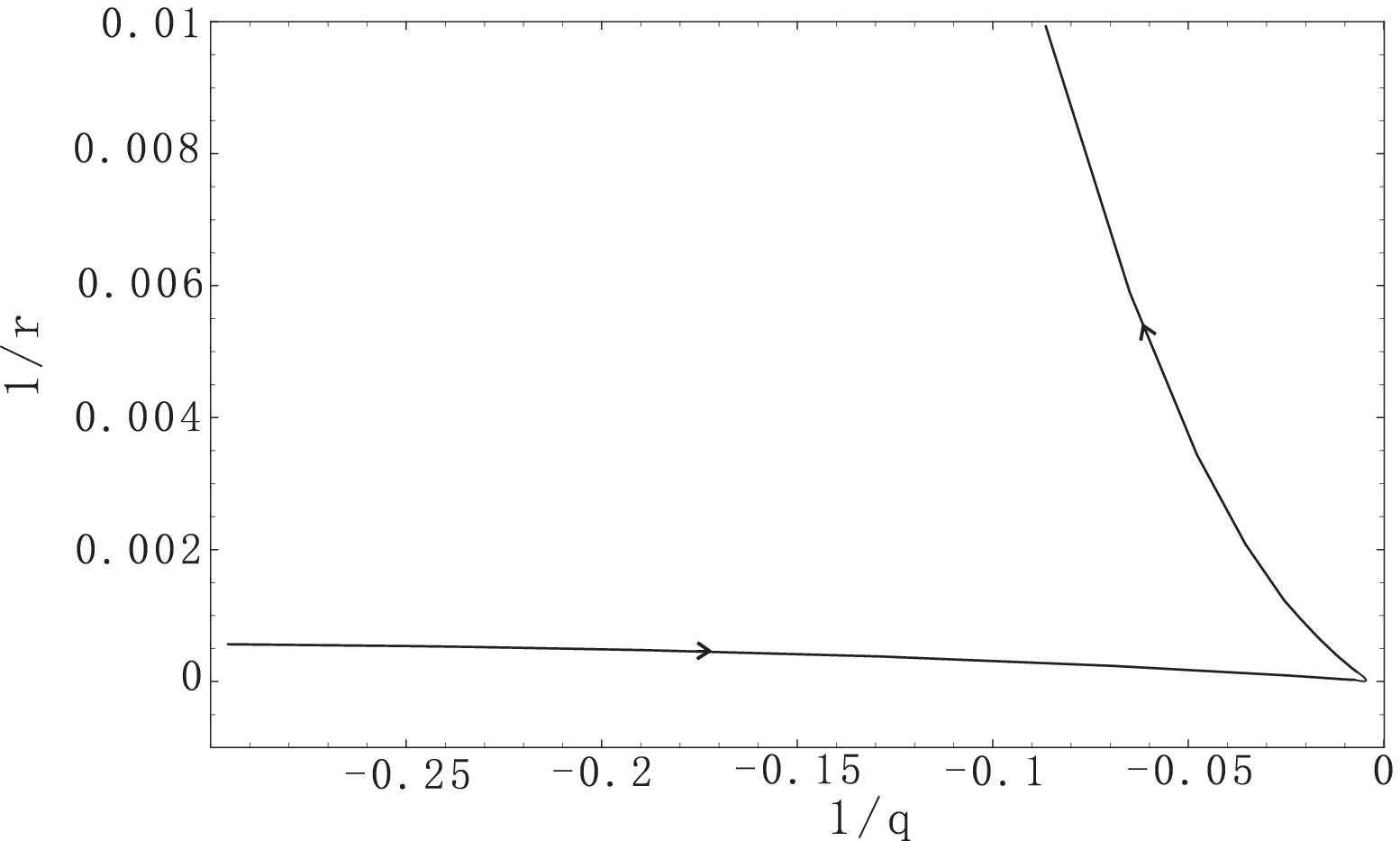}
\includegraphics[width=6.8cm]{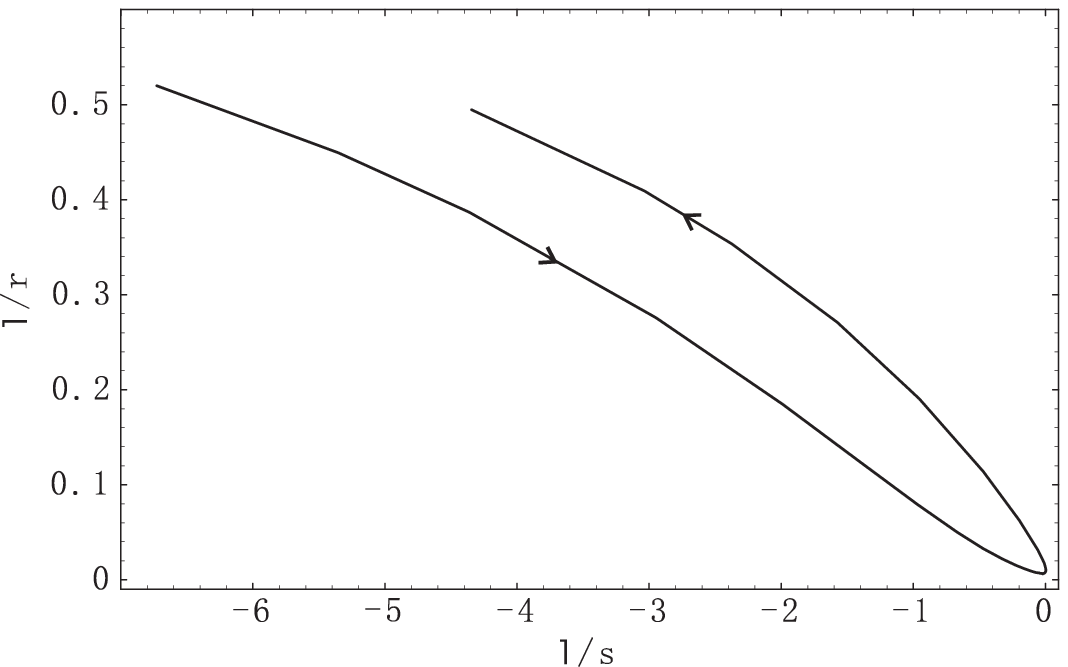}
\caption{Evolving trajectories of the statefinder parameters in the
$1/q-1/r$ and $1/s-1/r$ planes for the case of $-1\leq
a_{1}<-\frac{1}{9}$, where we choose $a_{1}=-\frac{1}{4}$ and $b=8$.
The arrows show the direction of the time evolution.}
 \label{planecase2}
\end{figure}
\begin{figure}[!htbp]
\includegraphics[width=7cm]{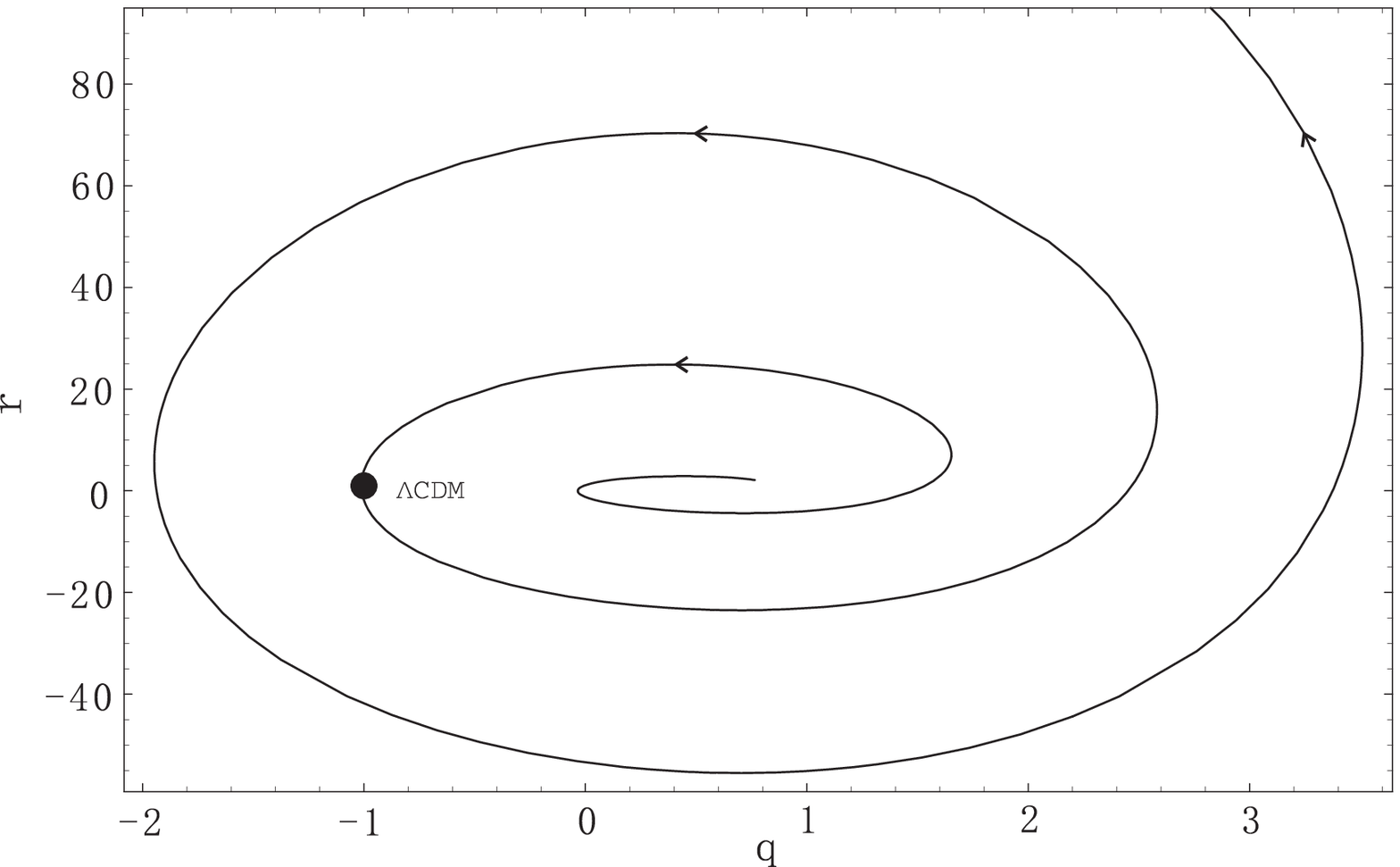}
\includegraphics[width=6.8cm]{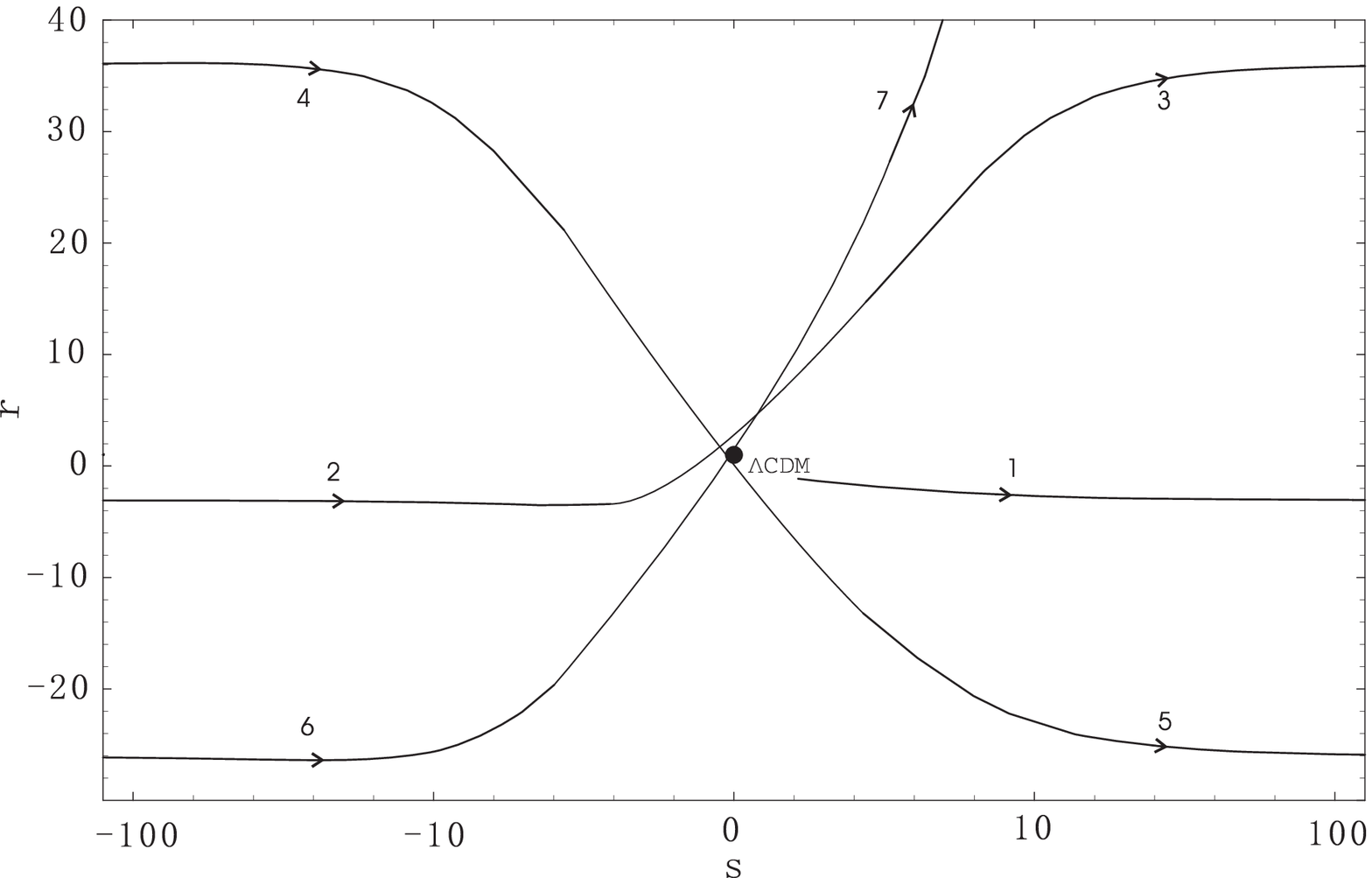}
\caption{ Evolving trajectories of the statefinder in the $q-r$ and
$s-r$ planes for the case of torsion parameter  $a_{1}<-1$, where we
choose the parameters $a_{1}=-2$ and $b=4$. The arrows show the
direction of the time evolution.}
 \label{planecase1}
\end{figure}
Thirdly, we discuss the time evolution of the trajectories for the
case of $-1 \leq a_1 < -\frac{1}{9}$. We plot evolving trajectories
in Fig. \ref{planecase2}, where we have chosen $a_1 = -\frac{1}{4}$
and $b = 8$. Obviously, the cosmic acceleration can happen since
deceleration parameter is negative. $|q|$, $|s|$ and $r$ become
larger and larger first, then less and less as the cosmic time
increase.

Finally, we consider the time evolution of the statefinder pairs
$\{r, s\}$ and $\{r, q\}$ in the case of $a_1 < -1$. In Fig.
\ref{planecase1}, we plot evolving trajectories in the $q-r$ and
$s-r$ planes, where we have chosen $a_1 = -2$ and $b = 4$. We find
easily that the evolving trajectories analogous to the case of $a_1
\geq 0$ except trajectories pass the $\Lambda$CDM point.

To sum up, it is very interesting to see that the scalar torsion
naturally provide the accelerating force in the universe for any
torsion parameter $a_1$. However, it is dependent on torsion
parameters that there is a decelerating ($q > 0$) expansion before
an accelerating ($q < 0$) expansion. The statefinder diagnostics
show that the universe naturally has an accelerating expansion at
low redshifts (late time) and a decelerating expansion at high
redshifts (early time) for the cases of $-\frac{1}{9} \leq a_1$ and
$a_1 < -1$. Obviously, scalar torsion cosmology can avoid some of
the problems which occur in other models. If we refuse the
possibility of non-positivity of the kinetic energy, we will employ
normal assumption, i. e., $a_1 > 0$. In this case, the effect of
torsion can make the expansion rate oscillate. With suitable
adjustments of the torsion parameters, it is possible to change the
quasi-period of the expansion rate as well as its amplitudes. It is
worth noting that the true values of the statefinder parameters of
the universe should be determined in model-independent way. In
principle, ${r_0, s_0}$ can be extracted from some future
astronomical observations, especially the SNAP-type experiment.

Why there are new features for the statefinder diagnostic of torsion
cosmology? Why the torsion parameter $a_1$ is divided into
differential ranges by the statefinder diagnostic? Answer is very
simple. In fact, the statefinder diagnostic has a direct bearing on
the attractor of cosmological dynamics. Therefore, we will discuss
the dynamic analysis in next section.

\section{Dynamics analysis}
Eqs. (\ref{dtH})-(\ref{dtR}) is an autonomous system, so we can use
the qualitative method of ordinary differential equations. Critical
points are always exact constant solutions in the context of
autonomous dynamical systems. These points are often the extreme
points of the orbits and therefore describe the asymptotic behavior.
If the solutions interpolate between critical points they can be
divided into a heteroclinic orbit and a homoclinic orbit (a closed
loop). The heteroclinic orbit connects two different critical points
and homoclinic orbit is an orbit connecting a critical point to
itself. In the dynamical analysis of cosmology, the heteroclinic
orbit is more interesting \cite{r8}. If the numerical calculation is
associated with the critical points, then we will find all kinds of
heteroclinic orbits.

According to equations (\ref{dtH})-(\ref{dtR}), we can obtain the
critical points and study the stability of these points.
Substituting linear perturbations $R=R_c+\delta R$,
$\Phi=\Phi_c+\delta\Phi$ and $H=H_c+\delta H$ near the critical
points into three independent equations, to the first orders in the
perturbations, gives the evolution of the linear perturbations, from
which we could yield three eigenvalues. Stability requires the real
part of all eigenvalues to be negative. There are five critical
points $(H_c, \Phi_c, R_c)$ of the system as follows
\begin{eqnarray}
&(\texttt{i}) & (0,0,0)\nonumber\\
&(\texttt{ii}) & \left(
\left(\frac{3(1+a_{1})}{8}-A\right)\sqrt{B+C},\frac{3}{2}\sqrt{B+C},-\frac{6(1+a_{1})}{b}\right)\nonumber\\
&(\texttt{iii}) & \left(-
\left(\frac{3(1+a_{1})}{8}-A\right)\sqrt{B+C},-\frac{3}{2}\sqrt{B+C},-\frac{6(1+a_{1})}{b}\right)\nonumber\\
&(\texttt{iv}) & \left(
\left(\frac{3(1+a_{1})}{8}+A\right)\sqrt{B-C},\frac{3}{2}\sqrt{B-C},-\frac{6(1+a_{1})}{b}\right)\nonumber\\
&(\texttt{v}) & \left(-
\left(\frac{3(1+a_{1})}{8}+A\right)\sqrt{B-C},-\frac{3}{2}\sqrt{B-C},-\frac{6(1+a_{1})}{b}\right)\nonumber\\\label{criticalpoints}
\end{eqnarray}
where $A=\frac{\sqrt{a_1^2(1+a_1)^3(1+9a_1)}}{8a_1(1+a_1)}$,
$B=-\frac{(1+a_1)(5+9a_1)}{a_1b}$ and
$C=-\frac{3\sqrt{a_1^2(1+a_1)^3(1+9a_1)}}{a_1^2b}$. The
corresponding eigenvalues of the critical points (i)-(v) are
\begin{eqnarray}
&&(\texttt{i})  (0,-\sqrt{-\frac{2(1+a_{1})}{a_{1}b}},\sqrt{-\frac{2(1+a_{1})}{a_{1}b}})\nonumber\\
&&(\texttt{ii})  \left(-\sqrt{B+C},
-3\left(\frac{3(1+a_{1})}{8}-A\right)\sqrt{B+C},
-\left(\frac{1+9a_{1}}{8}-3A\right)\sqrt{B+C}\right)\nonumber\\
&&(\texttt{iii})   \left(\sqrt{B+C},
3\left(\frac{3(1+a_{1})}{8}-A\right)\sqrt{B+C},\left(\frac{1+9a_{1}}{8}-3A\right)\sqrt{B+C}\right)\nonumber\\
&&(\texttt{iv})  \left(-\sqrt{B-C},
-3\left(\frac{3(1+a_{1})}{8}+A\right)\sqrt{B-C},-\left(\frac{1+9a_{1}}{8}+3A\right)\sqrt{B-C}\right)\nonumber\\
&&(\texttt{v})  \left(\sqrt{B-C},
3\left(\frac{3(1+a_{1})}{8}+A\right)\sqrt{B-C},\left(\frac{1+9a_{1}}{8}+3A\right)\sqrt{B-C}\right)\label{eigenvalues}
\end{eqnarray}

Using Eq. (\ref{criticalpoints}), we find that there is only a
critical point $(0, 0, 0)$ in the case of $a_1 \geq 0$. From Eq.
(\ref{eigenvalues}), the corresponding eigenvalue is $(0,
-\sqrt{\frac{2\mu}{a_1b}}i, \sqrt{\frac{2\mu}{a_1b}i})$, so $(0, 0,
0)$ is an asymptotically stable focus. If we consider the linearized
equations, then Eqs. (\ref{dtH})-(\ref{dtR}) are reduced to
\begin{eqnarray}
       \dot{H}=\frac{\mu}{6a_1}R,\qquad  \dot{\Phi}=\frac{1}{2a_1}R,\qquad
      \dot{R}=-\frac{4\mu}{b}\Phi\,.\label{dtHPHIRlinear}
\end{eqnarray}\\
The linearized system (\ref{dtHPHIRlinear}) has an exact periodic
solution
\begin{eqnarray}
       &&H=-\alpha R_{0}\sin \omega t+\frac{\mu}{3}\Phi_{0}\cos \omega t+H_{0}-\frac{\mu}{3}\Phi_{0},\nonumber\\
      &&\Phi=-\beta^{-1}R_{0}\sin \omega t +\Phi_{0}\cos \omega t\nonumber\\
      &&R=R_{0}\cos\ \omega t+\beta\Phi_{0}\sin \omega t\label{HRphips}
\end{eqnarray}
where $\omega = -\sqrt{\frac{2\mu}{|a_1|b}}$, $\alpha =
\sqrt{\frac{b\mu}{72|a_1|}}$, $\beta = \sqrt{\frac{8\mu |a_1|}{b}}$
and $H_0 = H(0)$, $\Phi_0 = \Phi(0)$ and $R_0 = R(0)$ are initial
values. Obviously, $(H, 0, 0)$ is a critical line of center for the
linearized Eqs. (\ref{HRphips}). In other words, there are only
exact periodic solutions for the linearized system, but there are
quasi-periodic solutions near the focus for the coupled nonlinear
equations. This property of quasi-periodic also appears in the
statefinder diagnostic with the case of $a_1 \geq 0$.
\begin{table}[thbp]
\caption{The physical properties of critical points for
$-\frac{1}{9}\leq a_{1}<0$} \label{cripointsa1l0}
\begin{tabular*}{\textwidth}{@{}l*{15}{@{\extracolsep{0pt plus
12pt}}l}} \br
critical points \qquad &property \qquad&$w_{eff}$ \qquad&stability \qquad\\
\mr
(\texttt{i})& saddle & $-\infty$& unstable\\
(\texttt{ii})&positive attractor&-1&stable\\
(\texttt{iii}) &negative attractor&-1&unstable\\
(\texttt{iv}) &saddle&-1&unstable\\
(\texttt{v}) &saddle&-1&unstable\\
\br
\end{tabular*}
\end{table}
\begin{table}[thbp]
\caption{The physical properties of critical points for $a_{1}<-1$}
\label{cripointsa1l-1}
\begin{tabular*}{\textwidth}{@{}l*{15}{@{\extracolsep{0pt plus
12pt}}l}} \br
critical points \qquad &property \qquad&$w_{eff}$ \qquad&stability \qquad\\
\mr
(\texttt{i})& focus & $\pm\infty$ & stable\\
(\texttt{ii})&saddle&-1&unstable\\
(\texttt{iii}) &saddle&-1&unstable\\
(\texttt{iv}) &saddle&-1&unstable\\
(\texttt{v}) &saddle&-1&unstable\\
\br
\end{tabular*}
\end{table}

\begin{figure}[!htbp]
\centering
\includegraphics[width=8.5cm]{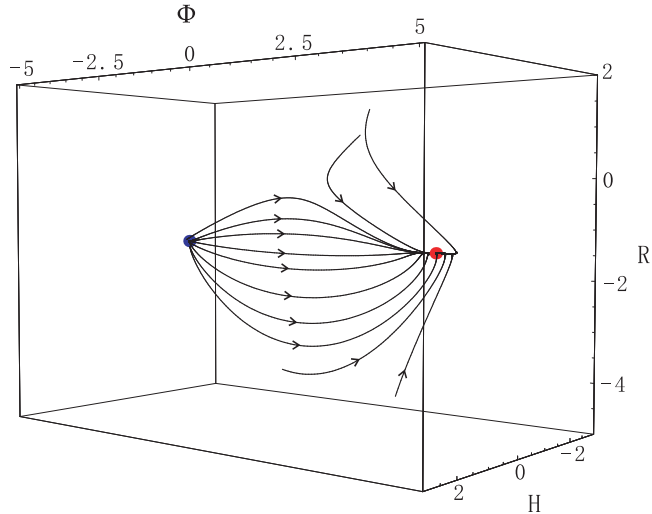}
\caption{The phase diagrams of $(H,\Phi,R)$ with $-1/9 \leq
a_{1}<0$. The heteroclinic orbit connects the critical points case
(iii) to case (ii). We take $a_{1}=-1/10,b=4$.}
 \label{heteroclinicorbit}
\end{figure}

In the case of $-1/9 \leq a_1 < 0$, the critical point (ii) is a
late time de Sitter attractor and (iii) is a negative attractor. The
properties of the critical points are shown in table
\ref{cripointsa1l0}. The de Sitter attractor indicates that torsion
cosmology is an elegant scheme and the scalar torsion mode is an
interesting geometric quantity for physics. In the dynamical
analysis of cosmology, the heteroclinic orbit is more interesting.
Using numerical calculation, we plot the heteroclinic orbit connects
the critical point case (iii) to case (ii) in Fig.
\ref{heteroclinicorbit}. This heteroclinic orbit is just
corresponding to the loop in Fig. \ref{planecase3}, which is from
$\Lambda$CDM point to $\Lambda$CDM point. Furthermore, the
trajectories with the shape of tadpole correspond to saddles.
\begin{figure}[!htbp]
\centering
\includegraphics[width=8.5cm]{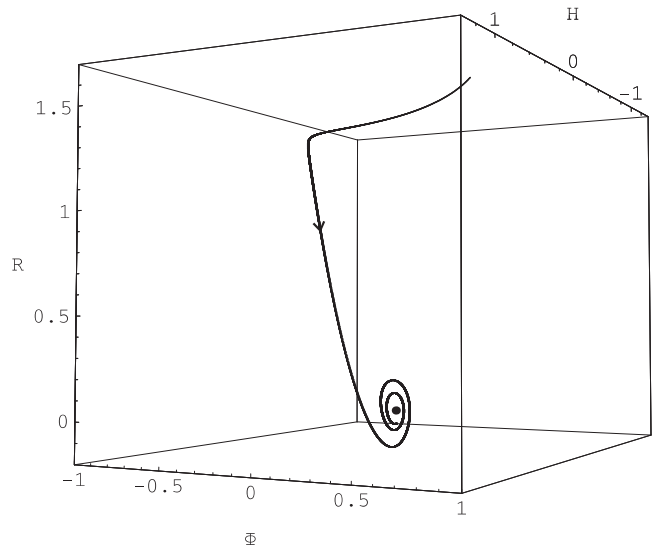}
\caption{The phase diagrams of $(H,\Phi,R)$ with $a_{1}<-1$. We take
$a_{1}=-2,b=4$ and the initial value $(5,1,\frac{29}{20})$.
$(0,0,0)$ is an asymptotically stable focus point.}
 \label{Focusa1l-1}
\end{figure}

In the case of $-1 \leq a_1 < -\frac{1}{9}$, there is only an
unstable saddle $(0, 0, 0)$ where the effective equation of state
$w_{eff}$ tends to $-\infty$. Therefore, the trajectories in  Fig.
\ref{planecase2} show that $|q|$,$|s|$ and $r$ become larger and
larger, then less and less as time increases.

In the case of $a_1 < -1$, the properties of the critical points are
shown in table \ref{cripointsa1l-1}. The trajectories correspond to
the stable focus (see Fig. 6) and unstable saddles with $w_{eff} =
-1$. Therefore, the trajectories pass through the $\Lambda$CDM
point.

\section{Fit the torsion parameters to supernovae date}
In Ref.\cite{r3}, the authors have compared the numerical values of
the torsion model with the observational data, in which they fixed
the initial values $H_0$, $\Phi_0$ and $R_0$, and torsion parameters
$a_1$ and $b$. In this section, we fixed the initial value, then fit
the torsion parameters to current type Ia supernovae data.

The scalar torsion cosmology predict a specific form of the Hubble
parameter $H(z)$ as a function of redshifts $z$ in terms of two
parameters $a_1$ and $b$ when we chose initial values. Using the
relation between $d_{L}(z)$ and the comoving distance $r(z)$ (where
$z$ is the redshift of light emission)
\begin{eqnarray}
d_{L}(z)=r(z)(1+z)\label{dis}
\end{eqnarray}
and the light ray geodesic equation in a flat universe
$cdt=a(z)dr(z)$ where $a(z)$ is the scale factor.

In general, the approach towards determining the expansion history
$H(z)$ is to assume an arbitrary ansatz for $H(z)$ which is not
necessarily physically motivated  but is specially designed to give
a good fit to the data for $d_L (z)$. Given a particular
cosmological model for $H(z; a_1, ... ,a_n)$ where $a_1, ...,a_n$
are model parameters, the maximum likelihood technique can be used
to determine the best fit values of parameters as well as the
goodness of the fit of the model to the data. The technique can be
summarized as follows: The observational data consist of $N$
apparent magnitudes $m_i (z_i)$ and redshifts $z_i$ with their
corresponding errors $\sigma_{m_i}$ and $\sigma_{z_i}$. These errors
are assumed to be gaussian and uncorrelated. Each apparent magnitude
$m_i$ is related to the corresponding luminosity distance $d_L$ by
\begin{equation}
\label{mz1}
 m(z)=M + 5 \; \log_{10} \left[{{d_L (z)}\over
{\mbox{Mpc}}}\right] + 25,
\end{equation}
where $M$ is the absolute magnitude. For the distant SNeIa, one can
directly observe their apparent magnitude $m$ and redshift $z$,
because the absolute magnitude $M$ of them is assumed to be
constant, i.e., the supernovae are standard candles. Obviously, the
luminosity distance $d_L (z)$ is the `meeting point' between the
observed apparent magnitude $m(z)$ and the theoretical prediction
$H(z)$. Usually, one define distance modulus $\nu(z)\equiv m(z)-M$
and express it in terms of the dimensionless `Hubble-constant free'
luminosity distance $D_L$ defined by$D_L (z) = {{H_0 d_L (z)}/ c}$
as
\begin{equation}
\label{mz2}
 \nu(z)=5\; \log_{10}(D_L (z))+\nu_0,
\end{equation}
where the zero offset $\nu_0$ depends on $H_0$ (or $h$) as
 \begin{equation}
\label{bm1} \nu_0 =5\; \log_{10} \left({{cH_0^{-1}}\over
{\mbox{Mpc}}}\right) + 25=-5\log_{10}h+42.38.
\end{equation}
The theoretically predicted value $D_L^{th} (z)$ in the context of a
given model $H(z;a_1,...,a_n)$ can be described by
\cite{starobinsky}
 \begin{equation}
 \label{dth1}
 D_L^{th} (z) = (1+z)
\int_0^z dz' \; {{H_0}\over {H(z';a_1,...a_n)}}.
\end{equation}
Therefore, the best fit values for the torsion parameters ($a_1, b$)
of the model are found by minimizing the quantity
 \begin{equation}
 \label{chi2def}
 \chi^2 (a_1, b)=\sum_{i=1}^N
 \frac{\left[\nu^{{obs}}(z_i)-5\log_{10}D_L^{{th}}(z_i;a_1, b)-\nu_0\right]^2}{\sigma_i^2}.
\end{equation}
Since the nuisance parameter $\nu_0$ is model-independent, its value
from a specific good fit can be used as consistency test of the data
\cite{r16} and one can choose \emph{a priori} value of it
(equivalently, the value of dimensionless Hubble parameter $h$) or
marginalize over it thus obtaining
\begin{equation}
\widetilde{\chi}^2(a_1, b)=A(a_1, b) -\frac{B(a_1,
b)^2}{C}+\ln\left(\frac{C}{2\pi}\right),
\end{equation}
where
\begin{equation}\label{A}
   A(a_1, b)=\sum_{i=1}^N
 \frac{\left[\nu^{{obs}}(z_i)-5\log_{10}D_L^{{th}}(z_i;a_1, b)\right]^2}{\sigma_i^2},
\end{equation}
\begin{equation}\label{B}
       B(a_1, b)=\sum_{i=1}^N
 \frac{\left[\nu^{{obs}}(z_i)-5\log_{10}D_L^{{th}}(z_i;a_1, b)\right]}{\sigma_i^2},
\end{equation}
and
\begin{equation}\label{C}
   C=\sum_{i=1}^N
 \frac{1}{\sigma_i^2}.
\end{equation}
In the latter approach, instead of minimizing ${\chi}^2(a_1, b)$,
one can minimize $\widetilde{\chi}^2(a_1, b)$ which is independent
of $\nu_0$.

The Eqs. (\ref{gradR}-\ref{torT}) can be solved explicitly by a
series in the form
\begin{equation}\label{E1}
   H(t,t_0)=H_0[1+\sum^{\infty}_{n=1}h_nu^{n}],
\end{equation}
where $u(t)=e^{\omega (t-t_0)}-1$ and
\begin{eqnarray}\label{E2}
 &&h_1=-\frac{2H_0}{\omega}-\frac{H_0}{2\omega a_1}+\frac{R_0}{6\omega
 H_0}+\frac{R_0}{6\omega a_1H_0}-\frac{bH_0R_0}{12\omega
 a_1}+\frac{bR_0^2}{144\omega a_1
 H_0} \nonumber\\
 &&+ \frac{\Phi_0}{3\omega}+\frac{\Phi_0}{3\omega a_1}+\frac{bR_0\Phi_0}{18\omega
 a_1}-\frac{\Phi_0^2}{18\omega H_0}-\frac{\Phi_0^2}{18\omega
 a_1H_0}-\frac{bR_0\Phi_0^2}{108\omega a_1H_0},\\
 &&h_2=\frac{H_0}{\omega}+\frac{H_0}{4\omega
 a_1}-\frac{2h_1H_0}{\omega }-\frac{h_1H_0}{2\omega a_1}-\frac{bh_1R_0H_0}{12\omega
 a_1}+\frac{bR_0^2}{288\omega a_1H_0}+\frac{h_1\Phi_0}{6\omega}\nonumber\\
 &&+\frac{h_1\Phi_0}{6\omega a_1}
 +\frac{bR_0\Phi_0}{36\omega a_1}+\frac{bh_1R_0\Phi_0}{36\omega
 a_1}-\frac{\Phi_0^2}{36\omega H_0}-\frac{\Phi_0^2}{36\omega a_1H_0}-\frac{bR_0\Phi_0^2}{108\omega
 a_1H_0},\nonumber\\
 \nonumber\\
 &&\cdots\cdots\cdots\cdots\nonumber
\end{eqnarray}
Using the general relation between Hubble parameter $H(t)$ and the
redshift $z$, $z$ can be written as a function of $t$
\begin{equation}\label{E3}
z(t)=\exp{[\int _{t}^{t_0}H(t)dt]}-1,
\end{equation}
However, the convergence radius of the series (\ref{E1}) is
$|\frac{\ln{2}}{\omega}|$, so we can use the expansion directly in
the case of the redshift being $z<z(t_0+\frac{\ln{2}}{\omega})$. By
the numerical calculation, we find that $t_*\sim
t_0+\frac{\ln{2}}{\omega}$ corresponds to $z_*\sim 0.45$ for the
valuses of parameters $R_0$ and $\Phi_0$ in the Fig.
\ref{192CLContours3}. For the case of $z>z_*$, we should use a
direct analytic continuation. Weierstrass \cite{Ahlfors} had built
the whole theory of analytic functions from the concept of power
series. Given a point $t_1=t_*+\alpha t_*$ ($0<\alpha <1$), the
function $H(t)$ has a Taylor expansion
\begin{equation}
H(t,t_1)=H(t_1,t_0)[1+\sum_{n=1}^{n}h_n(e^{\omega (t-t_1)}-1)^n].
\end{equation}
where the coefficients $h_n$ is still expressed as Eq. (\ref{E2})
and $H(t_1,t_0)$ can be decided by Eq. (\ref{E1}). The new series
defines an analytic function $H(t,t_1)$ which is said to be obtained
from $H(t,t_0)$ by direct analytic continuation. This process can be
repeated any number of times. In the general case we have to
consider a succession of power series $H(t,t_0)$,
$H(t,t_1)$,...,$H(t,t_m)$, each of which is a direct analytic
continuation of the preceding one. By using this method we have the
evolution of Hubble parameter $H(t)$. Furthermore, we have the
function $H(z)$ from Eq. (\ref{E3}). In fact, we need only to
consider the case of $z<2$ for the ESSENCE supernovae data.

 We now apply the above described maximum likelihood method
using the ESSENCE supernovae data which is one of the reliable
published data set consisting of 192 SNeIa ($N=192$). Beside the 162
data points given in table 9 of Ref. \cite{Wood}, which contains 60
ESSENCE SNeIa, 57 SNLS SNeIa and 45 nearby SNeIa, we add 30 SNeIa
detected at $0.216<z<1.755$ by the Hubble Space Telescope
\cite{Riess} as in Ref.\cite{Davis}.

\begin{figure}[!htbp]
\includegraphics[width=7cm]{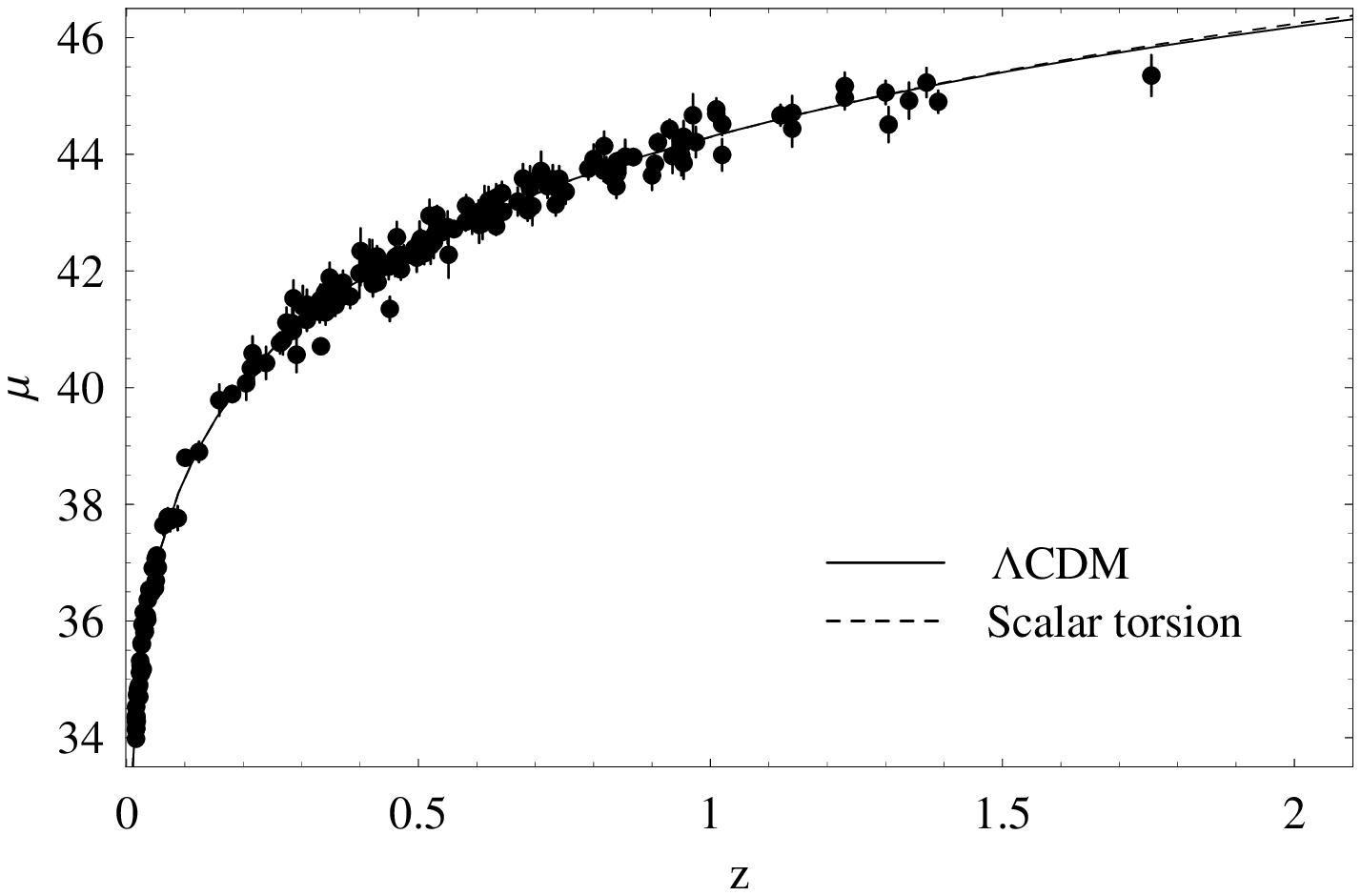}
\includegraphics[width=7.2cm]{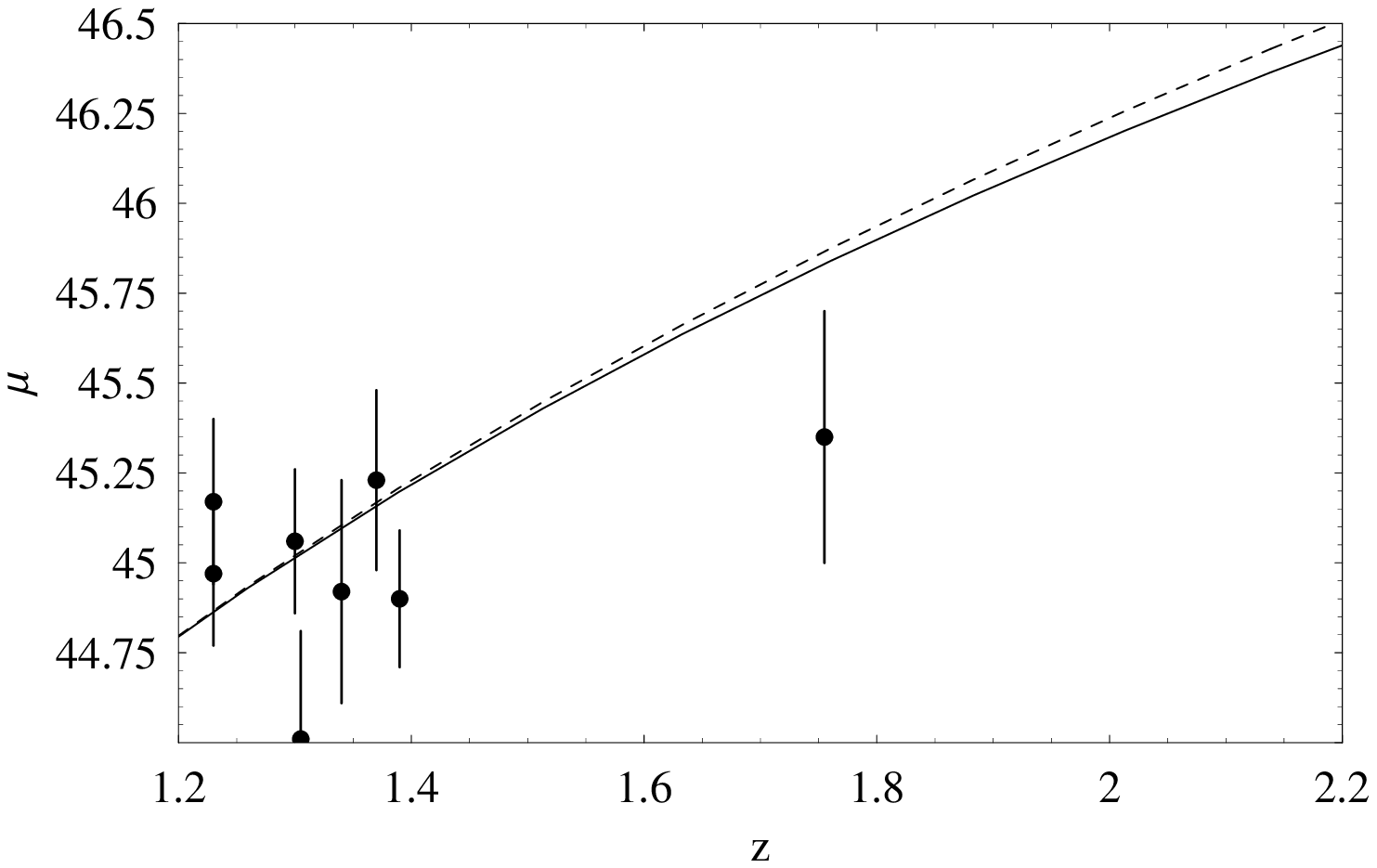}
\caption{Comparison of scalar torsion model and $\Lambda CDM$ with
the ESSENCE supernovae data via the relation between the redshift
$z$ and the distance modulus $\mu$. The scalar torsion
model($R_{0}=0.15$, $\Phi_{0}=0.34$, $a_{1}=-0.06$, $b=2.4$) gives a
close curve behavior to the one from $\Lambda CDM$
($\Omega_{m0}=0.26, \Omega_{\Lambda}=0.74$).}
 \label{zu}
\end{figure}

\begin{figure}[!htbp]
\centering
\includegraphics[width=8.5cm]{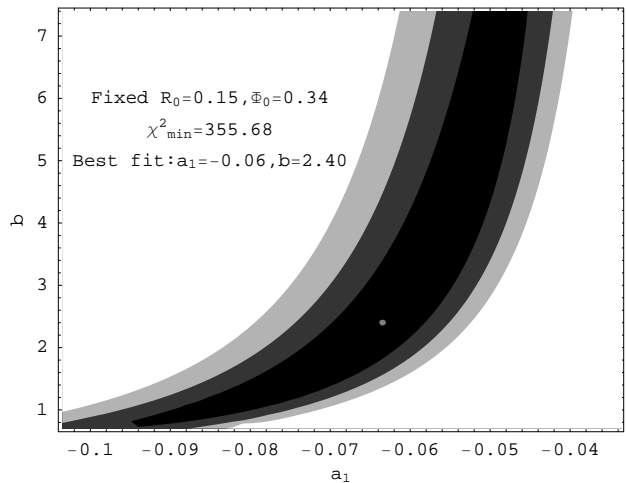}
\caption{The 68.3\%, 95.4\% and 99.7\% confidence contours of
torsion parameters $a_1$ and $b$ using the ESSENCE SNeIa dataset.
Here we have assumed $H_0=1$, $R_0=0.15$ and $\Phi_0=0.34$.}
\label{192CLContours3}
\end{figure}

In table \ref{differentvalues}, we show the best fit of torsion
parameters at different initial values of $R_0$ and $\Phi_0$.
\begin{table}[thbp]
\caption{The best fit of torsion parameters for different initial
values of $R_0$ and $\Phi_0$} \label{differentvalues}
\begin{tabular*}{\textwidth}{@{}l*{15}{@{\extracolsep{0pt plus
12pt}}l}} \br
 \qquad$R_0$ \qquad &$\Phi_0$ \qquad&$a_1$ \qquad&$b$\\
\mr
 \qquad0.25&0.35 &-0.10&1.44\\
 \qquad0.20&0.34 &-0.08&1.80\\
 \qquad0.15&0.34 &-0.06&2.40\\
 \qquad0.10&0.33 &-0.04&3.60\\
\br
\end{tabular*}
\end{table}
In Fig. \ref{192CLContours3}, contours with 68.3\%, 95.4\% and
99.7\% confidence level are plotted, in which we take a
marginalization over the model-independent parameter $\nu_0$. The
best fit as showed in the figure corresponds to $a_1=-0.06$ and
$b=2.40$, and the minimum value of $\chi^2=355.68$. For $\Lambda
CDM$, one can get $\chi^2(\Lambda CDM)=355.74$ and the best fit
$\Omega_{m0}=0.26$. Therefore, it's easy to know that $\Lambda CDM$
is consistent at the $1 \sigma$ level with the best fits of scalar
torsion cosmology. In Fig. \ref{zu}, we show a comparison of the
ESSENCE supernovae data along with the theoretically predicted
curves in the context of scalar torsion and $\Lambda CDM$ . We can
see that the scalar torsion model($R_{0}=0.15$, $\Phi_{0}=0.34$,
$a_{1}=-0.06$, $b=2.4$) gives a close curve behavior to the one from
$\Lambda CDM$ ($\Omega_{m0}=0.26, \Omega_{\Lambda}=0.74$). Clearly,
the allowed ranges of the parameters $a_1$ and $b$ favor the case of
$-\frac{1}{9} \leq a_1 < 0$ if we chose $R_0=0.15$ and
$\Phi_0=0.34$.

\section{Conclusion and discussion}
We have studied the statefinder diagnostic to the torsion cosmology,
in which an accounting for the accelerated universe is considered in
term of a Riemann-Cartan geometry: dynamic scalar torsion. We have
shown that statefinder diagnostic has a direct bearing on the
critical points. The statefinder diagnostic divides the torsion
parameter $a_1$ into four ranges, which is in keeping with the
requirement of dynamical analysis. Therefore, the statefinder
diagnostic can be used to an exceedingly general category of models
including several for which the notion of equation of state is not
directly applicable. The statefinder diagnostic has the advantage
over the dynamical analysis at the simplicity, but the latter can
provide more information.

The most interesting characteristic of the trajectories is that
there is a loop in the case of $-\frac{1}{9} \leq a_1 <0$. This
behavior corresponds to the heteroclinic orbit connecting the
negative attractor and de Sitter attractor. The trajectories with
the shape of tadpole show that they pass through the $\Lambda$CDM
fixed point along the time evolution, then the statefinder pairs are
going along with a loop and they will pass through the $\Lambda$CDM
fixed point again in the future. It is worth noting that there
exists closed loop in the Ref. \cite{r6}, but there is no closed
loop which contains the $\Lambda$CDM fixed point. These behaviors
indicate that torsion cosmology is an elegant scheme and the scalar
torsion mode is an interesting geometric quantity for physics.
Furthermore, the quasi-periodic feature of trajectories in the cases
of $a_1 \geq 0$ or $a_1 < -1$ shows that the numerical solutions in
Ref. \cite{r3} are not periodic, but are quasi-periodic near the
focus for the coupled nonlinear equations.

We fixed only the initial values, then fitted the torsion parameters
to current SNeIa dataset. We find that the scalar torsion naturally
explain the accelerating expansion of the universe for any torsion
parameter $a_1$. However, it is dependent $a_1$ and $b$ that there
is a decelerating expansion before an accelerating expansion. The
statefinder diagnostics show that the universe naturally have an
accelerating expansion at late time and a decelerating expansion at
early time for the case of $a_1 \geq -\frac{1}{9}$ and $a_1 < -1$.
If we refuse the possibility of non-positivity of the kinetic
energy, we have to employ normal assumption ($a_1 > 0$). Under this
assumption, the effect of torsion can make the expansion rate
oscillate. Furthermore, with suitable adjustments of the torsion
parameters and initial value, it is possible to change the
quasi-period of expansion rate as well as its amplitudes. In order
to have a quantitative understanding of the scalar torsion
cosmology, the matter density $\rho$, the effective mass density
$\rho_{eff}=\rho+\rho_T$, and the quantity $3p_{T}+\rho_{eff}$ are
important. This scenario bears a strong resemblance to the
braneworld cosmology in a very different context by Sahni, Shtanov
and Viznyuk \cite{Sahni}. The $\Omega_m$ parameters in the torsion
cosmology and in the $\Lambda$CDM cosmology can nevertheless be
quite different. Therefore, at high redshift, the torsion cosmology
asymptotically expands like a matter-dominated universe with the
value of $\Omega_m$ inferred from the observations of the local
matter density. At low redshift, the torsion model behaves like
$\Lambda$CDM but with a renormalized value of $\Omega_m^{\Lambda
CDM}$. The difference between $\Omega_m$ and $\Omega_m^{\Lambda
CDM}$ is dependent on the present value of statefinder parameters
{$r_0, s_0$}. A more detailed estimate, however, lies beyond the
scope of the present paper, and we will study it in a future work.
Finally, {$r_0$ and $s_0$} should be extracted from some future
astronomical observations in principle, especially the SNAP-type
experiments.

\section*{Acknowledgments} This work is supported by National
Science Foundation of China grant No. 10847153 and No. 10671128.


\section*{References}
\begin{thebibliography}{10}
\bibitem{r1} Sahni V and Starobinsky A 2000 {\it Int. J. Mod. Phys.}
             D{\bf9} 373 [astro-ph/9904398]\\
             Peebles P J E and Ratra B 2003 {\it Rev. Mod.
             Phys.} {\bf75} 559 [astro-ph/0207347]\\
             Padmanabhan T 2003 {\it Phys. Rept.} {\bf380} 235 [hep-th/0212290]\\
             Copeland E J, Sami M and Tsujikawa S 2006 {\it Int. J. Mod. Phys.}
             D{\bf15} 1753 [hep-th/0603057]\\
             Sahni V and Starobinsky A 2006 {\it Int. J. Mod. Phys.}
             D{\bf15} 2105 [astro-ph/0610026]
\bibitem{r2} Endo M and Fukui T 1977 {\it Gen. Rel. Grav.} {\bf8}
             833\\
             Overduin J M and Cooperstock F I 1998 {\it Phys. Rev.}
             D{\bf58} 043506 [astro-ph/9805260]\\
             Hao J G and Li X Z 2005 {\it Phys. Lett.} B{\bf606} 7 [astro-ph/0404154]\\
             Liu D J and Li X Z 2005 {\it Phys. Lett.} B{\bf611} 8 [astro-ph/0501596]\\
             Beesham A 1993 {\it Phys. Rev} D{\bf48} 3539
\bibitem{phantom}Caldwell R R 2002 {\it Phys. Lett.} B{\bf545} 23 [astro-ph/9908168]\\
             Sahni V and Shtanov Yu 2003 {\it JCAP} {\bf11} 014 [astro-ph/0202346]\\
             Hao J G and Li X Z 2003 {\it Phys. Rev.} D{\bf67}
             107303 [gr-qc/0302100]\\
             Liu D J and Li X Z 2003 {\it Phys. Rev.} D{\bf68}
             067301 [hep-th/0307239]\\
             Li X Z and Hao J G 2004 {\it Phys. Rev.} D{\bf69}
             107303 [hep-th/0303093]\\
             Hao J G and Li X Z 2003 {\it Phys. Rev.} D{\bf68}
             083514 [hep-th/0306033]
\bibitem{Kerlick} Kerlick G D 1976 {\it Ann. Phys.} \textbf{99}
                  127
\bibitem{Goenner} Goenner H and M\"{u}ller-Hoissen F 1984 {\it Class. Quant.
                  Grav.} \textbf{1} 651
\bibitem{Boeheretal} Capozziello S, Carloni S and Troisi A 2003 {\it Recent
                     Res. Dev. Astron. Astrophys.} \textbf{1} 625 [astro-ph/0303041]\\
                     Boehmer C G and Burnett J 2008 {\it Phys. Rev.} D{\bf78} 104001 [arXiv:0809.0469 [gr-qc]]\\
                     Boehmer C G 2005 {\it Acta Phys. Polon.} B\textbf{36} 2841 [gr-qc/0506033]\\
                     Mielke E W and Romero E S 2006 {\it Phys. Rev.} D{\bf73} 043521\\
                     Minkevich A V, Garkin A S and Kudin V I 2007 {\it Class. Quant. Grav.} {\bf24}
                     5835 [arXiv:0706.1157 [gr-qc]
\bibitem{r3} Shie K F, Nester J M and Yo H J 2008 {\it Phys. Rev.}
             D{\bf78} 023522 [arXiv:0805.3834 [gr-qc]]\\
             Yo H J and Nester J M 2007 {\it Mod. Phys. Lett.} A{\bf22} 2057 [astro-ph/0612738]\\
             Li X Z, Sun C B and Xi P 2009 {\it Phys. Rev.} D{\bf79}
             027301 [arXiv:0903.3088 [gr-qc]]
\bibitem{b1} Kopczy\'{n}ski W 1972 {\it Phys. Lett.} A{\bf39} 219
\bibitem{b2} Nester J M and Isenberg J A 1977 {\it Phys. Rev.}
             D{\bf15} 2078
\bibitem{b3} Hecht R D, Nester J M and Zhytnikov V V 1996 {\it Phys.
             Lett.} A{\bf222} 37\\
             Yo H J and Nester J M 1999 {\it Int. J. Mod. Phys.}
             D{\bf8} 459 [gr-qc/9902032]
\bibitem{b4} Shapiro I L 2002 {\it Phys. Rep.} {\bf357} 113 [hep-th/0103093]
\bibitem{r4} Hao J G and Li X Z 2004 {\it Phys. Rev.} D{\bf70}
             043529 [astro-ph/0309746]
\bibitem{r5} Sahni V, Saini T D, Starobinsky A A and Alam U 2003 {\it JETP
             Lett.} {\bf77} 201 [astro-ph/0201498]
\bibitem{r6} Alam U, Sahni V and Saini T D and Starobinsky A A 2003
             {\it Mon. Not. Roy. Astron. Soc.} {\bf344} 1057 [astro-ph/0303009]
\bibitem{r9} Zhang X 2007 {\it JCAP} {\bf0703} 007 [gr-qc/0611084]\\
             Chang B, Liu H, Xu L, Zhang C and Ping Y 2007 {\it
             JCAP} {\bf0701} 016 [astro-ph/0612616]\\
             Liu D J and Liu W Z 2008 {\it Phys. Rev.} D{\bf77}
             027301 [arXiv:0711.4854 [astro-ph]]
\bibitem{P}  Tretyakov P, Toporensky A, Shtanov Y and Sahni V 2006
             {\it Class. Quant. Grav.} {\bf23} 3259 [gr-qc/0510104]
\bibitem{F}  Hoyel F, Burbidge G and Narlikar J V 1993 {\it Astrophys.
             J.} {\bf410} 437 [astro-ph/9412045]
\bibitem{r7} Hehl F W, McCrea J D, Mielke E W and Neeman Y 1995 {\it Phys.
             Rept.} {\bf258} 1 [gr-qc/9402012]
\bibitem{r8} Li X Z, Zhao Y B and Sun C B 2005 {\it Class. Quant.
             Grav.} {\bf22} 3759 [astro-ph/0508019]
\bibitem{starobinsky} Starobinsky A A 1998 {\it JETP Lett.} {\bf68} 757 [astro-ph/9810431]\\
                      Liu D J, Sun C B and Li X Z 2006 {\it Phys. Lett.}
                      B{\bf634} 442 [astro-ph/0512355]
\bibitem{r16}Choudury T R and Padmanabhan T 2005 {\it Astron.
             Astrophys.} {\bf429} 807 [astro-ph/0311622]
\bibitem{Ahlfors} Ahlfors V 1979 {\it Complex analysis} (New York: McGraw Hill)
\bibitem{Wood} Wood-Vasey W M et al. 2007 {\it Astrophys. J} {\bf666}
               694 [astro-ph/0701041]
\bibitem{Riess} Riess A G et al. 2007 {\it Astrophys. J} {\bf659} 98
[astro-ph/0611572]
\bibitem{Davis} Davis T M et al. 2007 {\it Astrophys. J} {\bf666}
716 [astro-ph/0701510]
\bibitem{Sahni} Sahni V, Shtanov Y and Viznyuk A 2005 {\it J. Cosmol. Astropart.
                Phys.} {\bf0512} 005 [astro-ph/0505004]
\end{thebibliography}
\end{document}